\documentclass{PoS}
\usepackage{lineno}

\title{Searches for Lepton Number Violation and resonances in the $K^{\pm}\rightarrow \pi \mu \mu$ decays at the NA48/2 experiment}

\ShortTitle{Lepton Number Violation and resonances in the $K^{\pm}\rightarrow \pi \mu \mu$ decays}
\author{\speaker{Karim Massri}\thanks{On behalf of the NA48/2 Collaboration: G.~Anzivino, R.~Arcidiacono, W.~Baldini, S.~Balev, J.R.~Batley, M.~Behler, S.~Bifani, C.~Biino, A.~Bizzeti, B.~Bloch-Devaux
% ...
}\\
        University of Liverpool\\
        E-mail: \email{karim.massri@cern.ch}}

\abstract{The NA48/2 experiment at CERN collected a large sample of charged kaon decays into final states with multiple charged particles in 2003--2004. A new upper limit on the rate of the lepton number violating decay $\kpimmws$ obtained from this sample is reported: $\mathcal{B}(\kpimmws)<8.6 \times 10^{-11}$ at 90\% CL. Searches for two-body resonances in the $\kpimmns{\pm}$ decays (including heavy neutral leptons~$N_4$ and inflatons~$\chi$) in the accessible range of masses and lifetimes are also presented.
In the absence of a signal, upper limits are set on the products of branching ratios~$\mathcal{B}(\kmutwoN{\pm})\mathcal{B}(\Npimuns)$ and $\mathcal{B}(\kpichi{\pm})\mathcal{B}(\chimumu)$ as functions of the resonance mass and lifetime. These limits are in the $10^{-10}-10^{-9}$ range for resonance lifetimes below 100~ps.}

\FullConference{38th International Conference on High Energy Physics\\
		3-10 August 2016\\
		Chicago, USA}

\newcommand{\kpimmws}{K^{\pm} \to \pi^{\mp} \mu^{\pm} \mu^{\pm}}
\newcommand{\kpimmns}[1]{K^{#1} \to \pi \mu \mu}
\newcommand{\kpimm}[1]{K^{#1} \to \pi^{#1} \mu^{+} \mu^{-}}

\newcommand{\kmutwoN}[1]{K^{#1} \to \mu^{#1} N_{4}}
\newcommand{\kpichi}[1]{K^{#1} \to \pi^{#1} \chi}
\newcommand{\Npimuns}{N_{4} \to \pi\mu}
\newcommand{\Npimuws}{N_{4} \to \pi^{\mp}\mu^{\pm}}
\newcommand{\Npimurs}{N_{4} \to \pi^{\pm}\mu^{\mp}}
\newcommand{\chimumu}{\chi \to \mu^{+}\mu^{-}}

\newcommand{\kthreepic}[1]{K^{#1} \to \pi^{#1} \pi^+ \pi^-}
\newcommand{\kthreepin}[1]{K^{#1} \to \pi^{#1} \pi^0 \pi^0}
\newcommand{\kpmm}{K_{\pi\mu\mu}}
\newcommand{\kpmmlnv}{K_{\pi\mu\mu}^{\rm LNV}}
\newcommand{\kpmmlnc}{K_{\pi\mu\mu}^{\rm LNC}}
\newcommand{\npmmlnv}{N_{\pi\mu\mu}^{\rm LNV}}

\begin{document}
\section{Introduction}
The NA48/2 experiment at CERN SPS was a multi-purpose $K^{\pm}$ experiment which collected data in 2003--2004, whose main goal was to search for direct CP violation in the $\kthreepic{\pm}$ and $\kthreepin{\pm}$ decays~\cite{ba07}. The large statistics of the samples of charged kaon decays into final states with multiple charged particles collected allows to search for the forbidden LNV $\kpimmws$ decay, as well as for two-body resonances in $\kpimmns{\pm}$ decays.
Since a particle~$X$ produced in a $K^{\pm}\to\mu^{\pm}X$ ($K^{\pm}\to\pi^{\pm}X$) decay and decaying promptly to $\pi^{\pm}\mu^{\mp}$ ($\mu^+\mu^-$) would produce a narrow spike in the invariant mass $M_{\pi\mu}$ ($M_{\mu\mu}$) spectrum, the invariant mass distributions of the collected $\kpimmns{\pm}$ samples have been scanned looking for such a signature.

\boldmath
\section{Selected data samples}
\unboldmath
\label{sec:datasamples}
The event selection is based on the reconstruction of a three-track vertex: given the resolution of the vertex longitudinal position ($\sigma_{vtx} = 50$~cm), $\kpimmws$ and $\kpimm{\pm}$ decays (denoted $\kpmmlnv$ and $\kpmmlnc$ below) mediated by a short-lived ($\tau\lesssim 10$~ps) resonant particle are indistinguishable from a genuine three-track decay.
The size of the selected $\kpmm$ samples is normalised relative to the abundant $K^\pm\to\pi^\pm\pi^+\pi^-$ channel (denoted $K_{3\pi}$ below),
from which the number of $K^\pm$ decays in the 98~m long fiducial decay region is obtained: $N_K = (1.64\pm0.01)\times 10^{11}$.
The $\kpmm$ and $K_{3\pi}$ samples are collected concurrently using the same trigger logic.

\begin{figure}[h]
\begin{minipage}{0.5\textwidth}
\includegraphics[width=\textwidth]{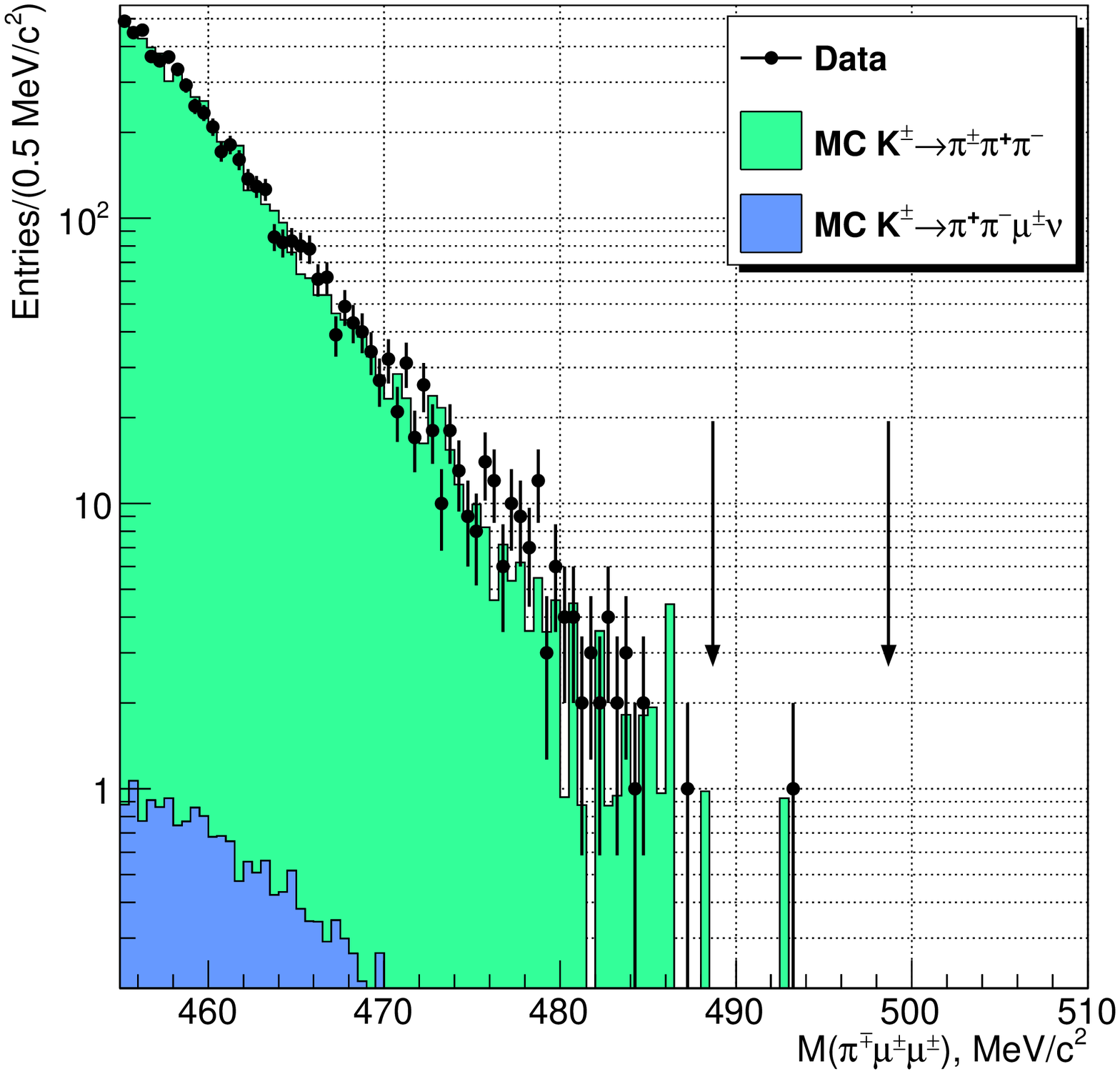}%
\end{minipage}
\hfill
\begin{minipage}{0.5\textwidth}
\includegraphics[width=\textwidth]{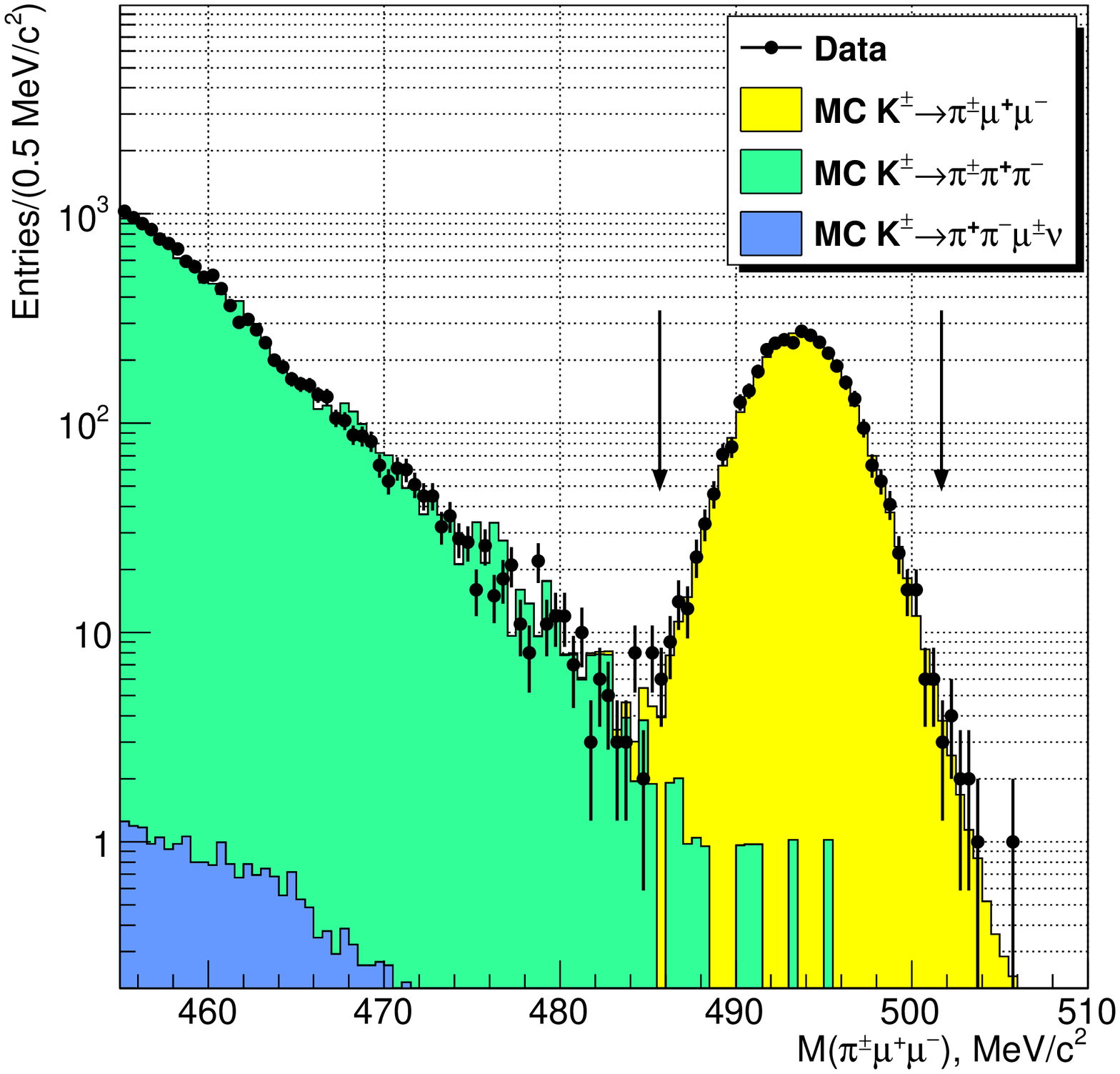}%
\end{minipage}
\caption{Invariant mass distributions of data and MC events passing the $\kpmmlnv$ (left) and
$\kpmmlnc$~(right) selections. The signal mass regions are indicated with vertical arrows.}\label{fig:mpimm}  
\end{figure}

The invariant mass distributions of data and MC events passing the $\kpmmlnv$ and $\kpmmlnc$ selections are shown in Fig.~\ref{fig:mpimm}.
One event is observed in the signal region after applying the $\kpmmlnv$ selection, while 3489 $\kpmmlnc$ candidates are selected with the $\kpmmlnc$ selection. A peak search assuming different mass hypotheses is performed over the distributions of the invariant masses $M_{ij}$ ($ij = \pi^{\pm}\mu^{\mp}, \mu^+\mu^-$) of the selected $\kpmm$ samples. In total, 284 (267) and 280 mass hypotheses are tested respectively for the search of resonances in the $M_{\pi\mu}$ distribution of the $\kpmmlnv$ ($\kpmmlnc$) candidates and in the $M_{\mu\mu}$ distribution of the $\kpmmlnc$ candidates, covering the full kinematic ranges.

\boldmath
\section{Results}
\unboldmath
\label{sec:results}
\subsection{Upper Limit on $\mathcal{B}(\kpimmws)$}
The upper limit~(UL) at 90\% confidence level~(CL) on the number of $\kpimmws$ signal events in the $\kpmmlnv$ sample corresponding
to the observation of one data event and a total number of expected background events $N_{bkg} = 1.163 \!\pm\! 0.867_{stat} \!\pm\! 0.021_{ext} \!\pm\! 0.116_{syst}$ is obtained applying an extension of the Rolke-Lopez method~\cite{ro01}: $\npmmlnv < 2.92$ at 90\%~CL. 
Using the values of the signal acceptance~$A(\kpmmlnv)=20.62\%$ estimated with MC simulations and the number~$N_K$ of kaon decays in the fiducial volume~(Sec.~\ref{sec:datasamples}), the UL on the number of $\kpimmws$ signal events in the $\kpmmlnv$ sample leads to a constraint on the signal branching ratio~$\mathcal{B}(\kpimmws)$:
\begin{equation}
\label{eq:BR_kpimmws_experimental}
\mathcal{B}(\kpimmws) = \frac{\npmmlnv}{N_K\cdot A(\kpmmlnv)}< 8.6 \times 10^{-11} \quad \mbox{@ 90\% CL}.
\end{equation}

\subsection{Results of the search for two-body resonances}
\label{subsubsec:n4mscan_ul}
\begin{figure}[p]
\begin{center}
\begin{minipage}{0.5\textwidth}
\includegraphics[width=\textwidth]{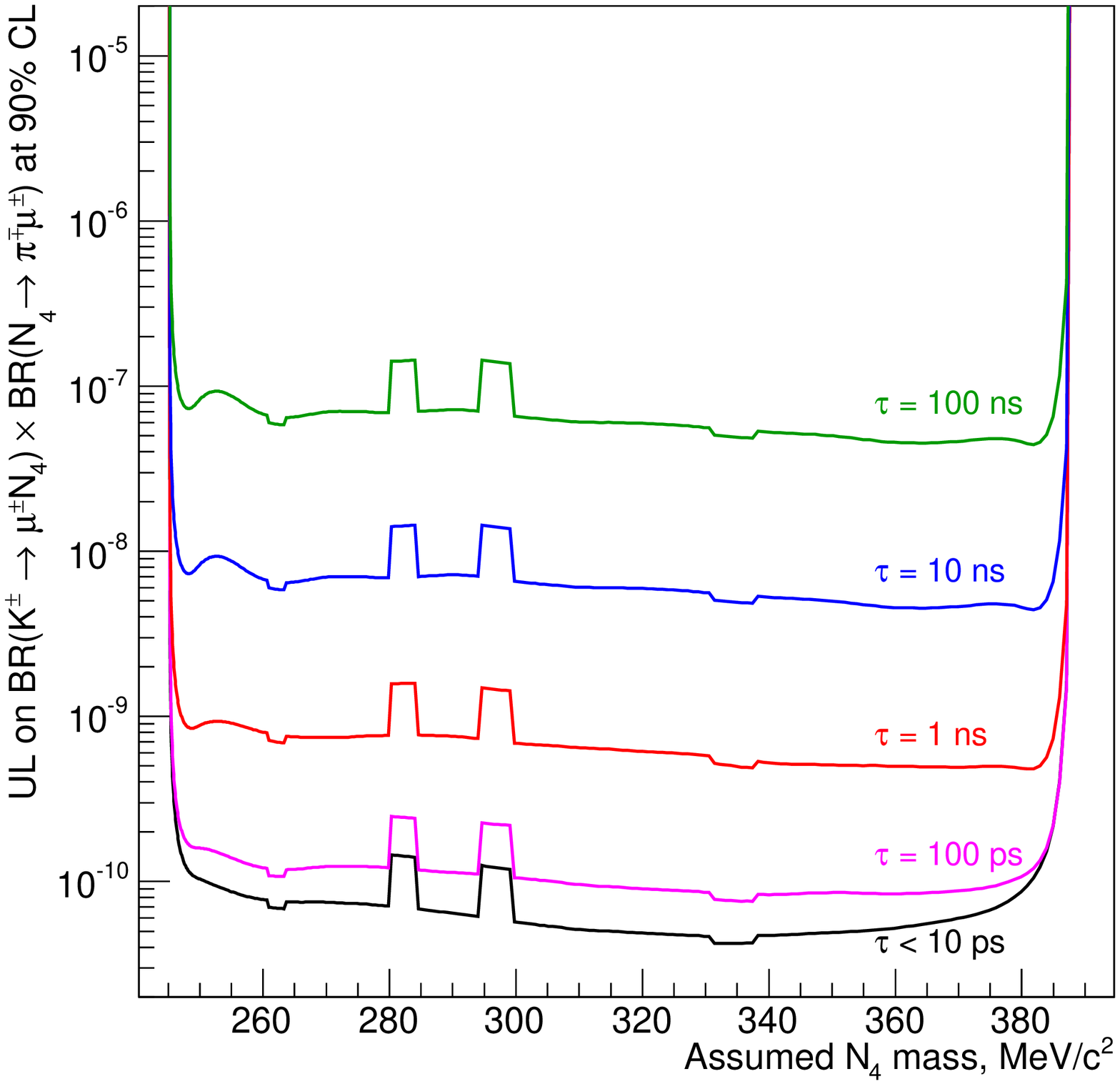}%
\end{minipage}
\put(-23,90){\Large\bf a}
\hfill
\begin{minipage}{0.5\textwidth}
\includegraphics[width=\textwidth]{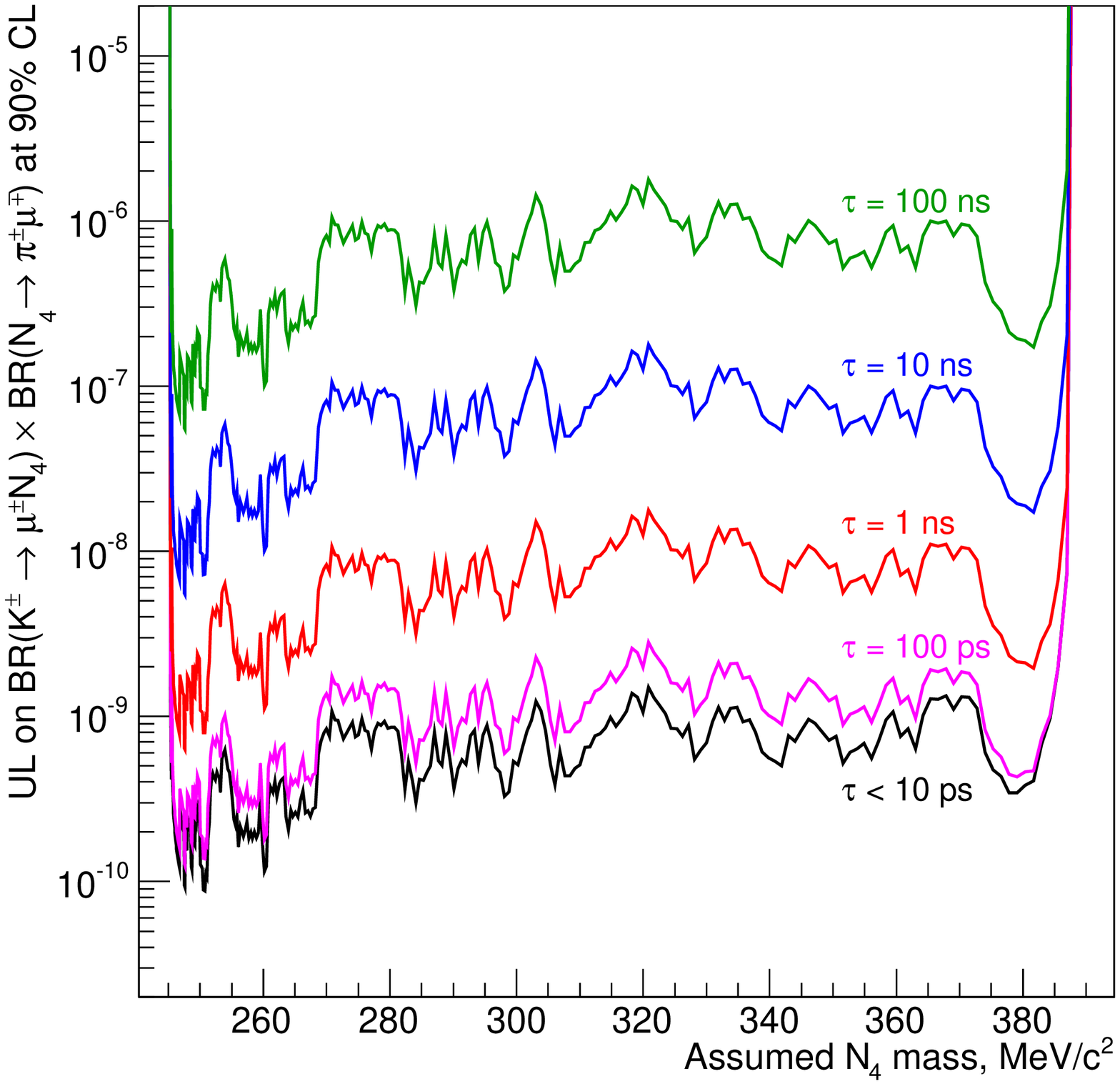}%
\end{minipage}
\put(-23,90){\Large\bf b}\\
\begin{minipage}{0.5\textwidth}
\includegraphics[width=\textwidth]{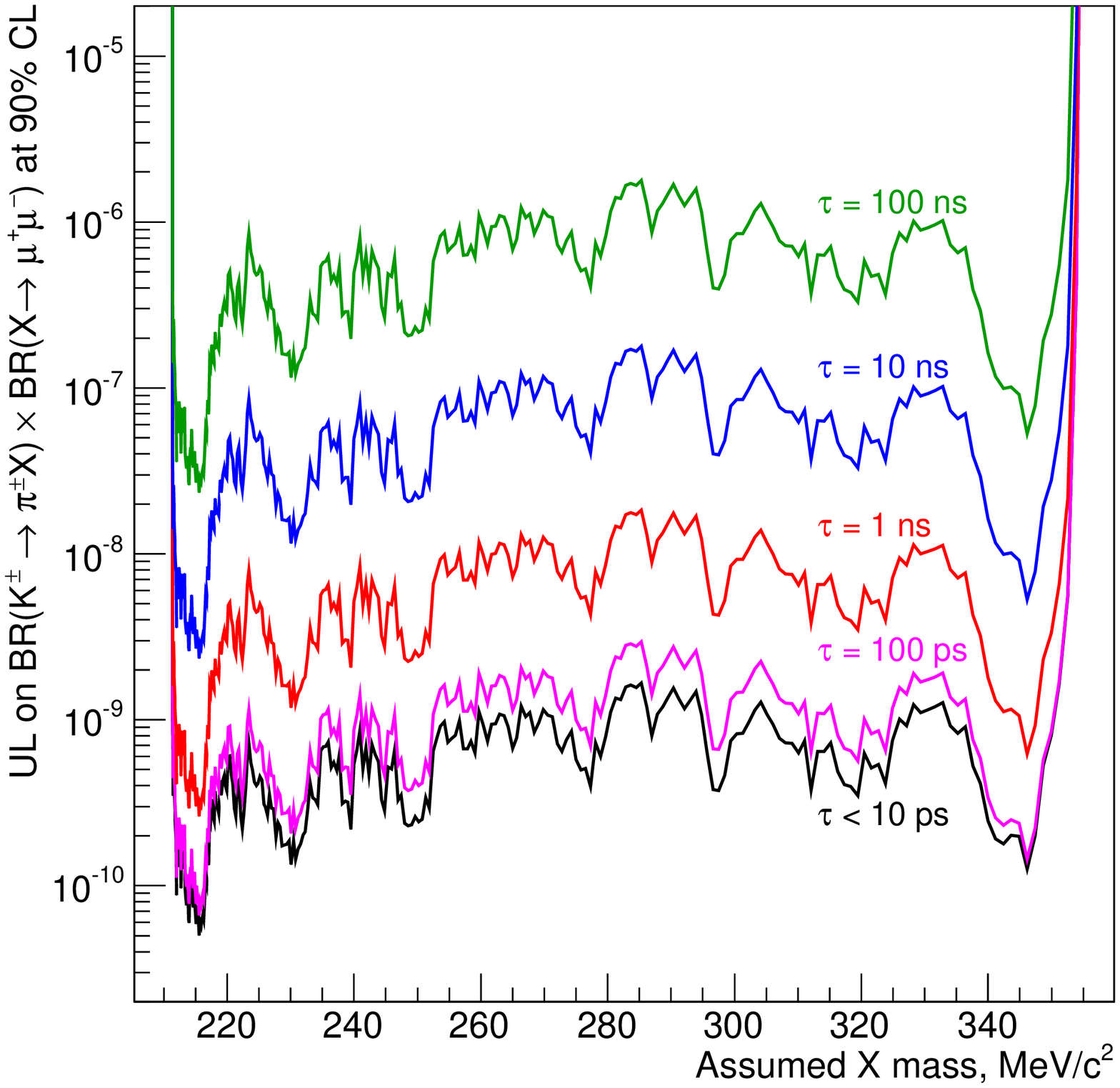}%
\end{minipage}
\put(-23,90){\Large\bf c}
\caption{Obtained ULs at 90\% CL on the products of branching ratios as functions of the resonance mass and lifetime: a)~$\mathcal{B}(\kmutwoN{\pm})\mathcal{B}(\Npimuws)$; b)~$\mathcal{B}(\kmutwoN{\pm})\mathcal{B}(\Npimurs)$; c)~$\mathcal{B}(\kpichi{\pm})\mathcal{B}(\chimumu)$.
All presented quantities are strongly correlated for neighbouring resonance masses as the mass step of the scan is about 8 times smaller than the signal window width.}\label{fig:kpimmws_results_data}  
\end{center}
\end{figure}

No signal is observed, as the local significances of the signals in each mass hypothesis never exceed 3 standard deviations.
In absence of a signal, ULs on the product~$\mathcal{B}(K^{\pm}\to p_1 X)\mathcal{B}(X\to p_2 p_3)$ ($p_1p_2p_3 = \mu^{\pm}\pi^{\mp}\mu^{\pm},\mu^{\pm}\pi^{\pm}\mu^{\mp},\pi^{\pm}\mu^{+}\mu^{-}$) as a function of the resonance lifetime~$\tau$ are obtained for each mass hypothesis~$m_i$, by using the values of the acceptances~$A_{\pi\mu\mu}(m_i,\tau)$ and the ULs on the number~$N^i_{sig}$ of signal events for such a mass hypothesis:
\begin{equation}
\label{eq:BR_res_massbin}
\left.\mathcal{B}(K^{\pm}\to p_1 X)\mathcal{B}(X\to p_2 p_3)\right|_{m_i,\tau} = \frac{N_{sig}^i}{N_K \cdot A_{\pi\mu\mu}(m_i,\tau)}.
\end{equation}
The obtained ULs on~$\mathcal{B}(K^{\pm}\to p_1 X)\mathcal{B}(X\to p_2 p_3)$ ($p_1p_2p_3 = \mu^{\pm}\pi^{\mp}\mu^{\pm},\mu^{\pm}\pi^{\pm}\mu^{\mp},\pi^{\pm}\mu^{+}\mu^{-}$) as a function of the resonance mass, for several values of the resonance lifetime, are shown in Fig.~\ref{fig:kpimmws_results_data}.

\section{Conclusions}
The searches for the LNV $\kpimmws$ decay and resonances in $\kpimmns{\pm}$ decays at the NA48/2 experiment, using the 2003--2004 data, are presented. 
No signals are observed. An UL of $8.6\times10^{-11}$ for $\mathcal{B}(\kpimmws)$ has been established, which improves the best previous limit~\cite{ba11} by more than one order of magnitude.
ULs are set on the products~$\mathcal{B}(\kmutwoN{\pm})\mathcal{B}(\Npimuns)$ and $\mathcal{B}(\kpichi{\pm})\mathcal{B}(\chimumu)$ as functions of the resonance mass and lifetime. These limits are in the $10^{-10}-10^{-9}$ range for resonance lifetimes below 100~ps.

\end{document}